\documentclass[reprint, superscriptaddress, amsmath,amssymb, aps, prl]{revtex4-1}

\usepackage{graphicx}  
\usepackage{dcolumn}   
\usepackage{bm}        
\usepackage{verbatim}  

\usepackage{physics}
\usepackage{float}
\usepackage{color}
\usepackage[dvipsnames]{xcolor}
\usepackage[colorlinks]{hyperref}

\usepackage{algorithm}
\usepackage[noend]{algpseudocode}

\def \ham{\mathcal{H}}

\setcitestyle{numbers,square}

\begin{document}

\title{Reinforcement Learning for Digital Quantum Simulation}
\author{Adrien Bolens and Markus Heyl}
\affiliation{Max-Planck-Institut  f{\"u}r  Physik  komplexer  Systeme,  N{\"o}thnitzer  Stra{\ss}e  38,  01187  Dresden,  Germany}

\date{\today}
\begin{abstract}
Digital quantum simulation is a promising application for quantum computers. Their free programmability provides the potential to simulate the unitary evolution of any many-body Hamiltonian with bounded spectrum by discretizing the time evolution operator through a sequence of elementary quantum gates, typically achieved using Trotterization. A fundamental challenge in this context originates from experimental imperfections for the involved quantum gates, which critically limits the number of attainable gates within a reasonable accuracy and therefore the achievable system sizes and simulation times. In this work, we introduce a reinforcement learning algorithm to systematically build optimized quantum circuits for digital quantum simulation upon imposing a strong constraint on the number of allowed quantum gates.
With this we consistently obtain quantum circuits that reproduce physical observables with as little as three entangling gates for long times and large system sizes. As concrete examples we apply our formalism to a long range Ising chain and the lattice Schwinger model. 
Our method makes larger scale digital quantum simulation possible within the scope of current experimental technology.
\end{abstract}

\maketitle

\paragraph{Introduction.} Digital quantum simulation (DQS) has emerged as one of the most promising applications of quantum computers. Unlike analog simulators, which directly mimic the Hamiltonian of interest, digital simulators reproduce a target time-evolution operator with a sequence of elementary quantum gates. In principle, the unitary time-evolution of any spin-type Hamiltonian can be encoded in a quantum computer with arbitrary precision \cite{feynman1999simulating}.
The experimental implementation of DQS has seen remarkable progress in the recent years leading to the simulation of theoretical condensed matter models \cite{barreiro2011open,lanyon2011universal,barends2015digital,salathe2015digital,langford2017experimentally,wei2018exploring}, lattice gauge theories \cite{martinez2016real}, and quantum chemistry problems \cite{o2016scalable,kandala2017hardware,hempel2018quantum}. A common and natural approach to factorize time evolution operators into elementary quantum gates is to utilise Suzuki-Trotter formulas \cite{trotter1959product,suzuki1976relationship}.
While the theoretical Trotter error can be well controlled \cite{lloyd1996universal,heyl2019quantum,sieberer2019digital}, high accuracy Trotterization requires a large number of quantum gates. This leads to a critical problem because each of these individual gates suffers from experimental imperfections, in particular those which entangle qubits. A key challenge of DQSs is therefore to identify factorizations of time evolution operators utilizing a minimal number of quantum gates in order to exploit currently available hardware resources optimally.

In this work, we introduce a method based on reinforcement learning (RL) to systematically build DQSs constrained to a fixed low number of entangling gates.
As a key step in our RL algorithm towards feasible large-scale DQS we propose to optimise the quantum circuits not with respect to the conventionally used global many-body wave function, but rather based on a local reward with the goal to reproduce expectation values of local observables and correlation functions. Remarkably, we find that the dynamics of strongly correlated systems can be digitally realised using just a handful of gates making large system sizes and long-time simulations feasible on current day devices. Specifically, for the lattice Schwinger model, we build quantum circuits using only three entangling gates that correctly reproduce the dynamics of local observables and correlation functions for up to $16$ qubits and for large times, reducing the number of entangling gates by one order of magnitude in comparison to a recent pioneering DQS experiment for $4$ qubits~\cite{martinez2016real}.
With our RL algorithm we are able to systematically build DQSs with a drastically reduced number of quantum gates for large quantum many-body systems pushing the design of quantum circuits beyond what has been achieved previously utilizing RL methods~\cite{fosel2018reinforcement,bukov2018reinforcement,bukov2018reinforcementPRB,yao2020policy} or in the field of quantum control~\cite{martinez2016real, kokail2019self}.
Our work provides a route towards larger-scale DQS in previously inaccessible regimes with currently available hardware resources.\\

\paragraph{Digital Quantum Simulation.}
Let $\ham = \sum_l \ham_l$ be such that $U_l(t)=\exp(-i\ham_l t)$ can be realized on the chosen quantum computing platform. The targeted dynamics can then be approximately factorised using the Suzuki-Trotter formula:
$
  e^{-i \ham \tau} \approx \left( \prod_l e^{-i \ham_l \tau/n} \right)^n
$
splitting the time $\tau$ of the simulation into $n$ smaller steps of duration $\tau/n$.
This Trotterization comes with an error that is rigorously upper bounded as $O(N\tau^2/n)$~\cite{lloyd1996universal} with $N$ the number of qubits whereas the error on local observables can be even much smaller~\cite{heyl2019quantum}.
The central problem is that higher Trotterization accuracy requires larger $n$.
This, however, increases the number of required quantum gates and therefore amplifies the imperfections due to faulty gate operations.
In this work we aim to generate optimized quantum circuits for the factorization of time-evolution operators with a minimal number of quantum gates. We focus on trapped ion quantum computing platforms with the following set of universal quantum gates consisting of the single-qubit rotations and the entangling gate,
\begin{align}
    U_j^x(\theta) = e^{-i \theta \sigma^x_j},
    U_j^z(\theta) = e^{-i \theta \sigma^z_j}, 
    U^{xx}(\theta) &= e^{-i \theta \sum\limits_{j<k} \frac{\sigma^x_j \sigma^x_k}{|k - j|^{\alpha}}}
    \label{eq:quantum_gates}
\end{align}
where $\sigma^{x}_j, \sigma^{y}_j$, and $\sigma^{z}_j$ are the Pauli matrices at site $j$.
The exponent $\alpha$ can be theoretically tuned within the range $0\leq \alpha <3$, but the optimal performance is typically reached either for $\alpha=0$ or $\alpha \approx 1$.
For the following we will focus for concreteness on either $\alpha=3$ or $\alpha=1$ while emphasizing that our approach can be straighforwardly applied also to other $\alpha$ or other quantum computing architectures such as superconducting qubits with different sets of universal quantum gates.

\begin{figure}[t]
\centering
\includegraphics[width=0.45\textwidth]{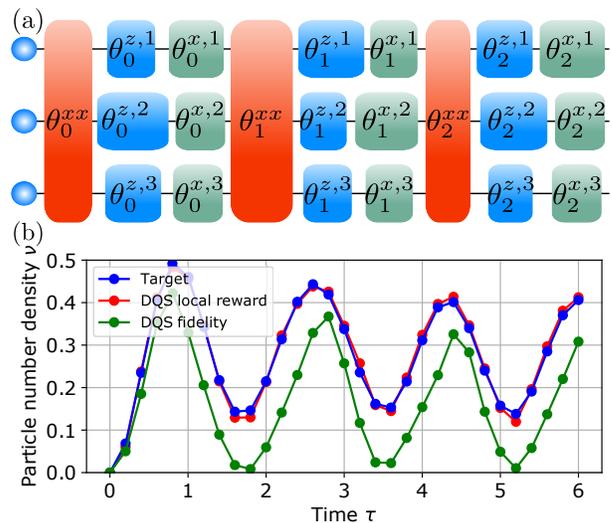}
    \caption{\small (a) Generic quantum circuit used for the DQSs of the long-range Ising and Schwinger models, here for a three-site system. The boxes represent different quantum gates, each with a different variational parameter depicted by the box width. (b) Particle number density $\nu$ in the 10-site lattice Schwinger model starting from the bare vacuum for the parameters used in a recent DQS experiment~\cite{martinez2016real}: $J=w=2m$ in Eq.~\eqref{eq:schwinger}. We show the DQS results using the fidelity (green) and the local reward~\eqref{eq:local_reward} (red), and the exact time evolution (blue).}
	\label{fig:architecture}
\end{figure}

The central goal of our work is to find circuits with a small number of quantum gates for the task of reproducing the dynamics of a given Hamiltonian. 
We translate this task into a variational optimization problem as follows.
Let $|\psi_0\rangle$ denote the initial state and let us fix the resources in terms of quantum gates as in Eq.~\eqref{eq:quantum_gates}.
Then we construct a sequence of gates:
\begin{align}
  |\psi_{\rm DQS}\rangle &= U_{n} \cdots U_2 U_1 |\psi_0\rangle \, , \\
  U_t &= U^{xx}(\theta^{xx}_t) \prod_j [U^z_j(\theta^{z, j}_t) U^x_j(\theta^{x, j}_t)] \, ,
  \label{eq:action_step}
\end{align}
as depicted schematically in Fig.~\ref{fig:architecture} (a).
The main goal now is to choose the underlying variational parameters $\theta = (\theta^{xx}_t, \theta_t^{z, 1}, \theta_t^{x, 1}, \dots, \theta_t^{z, N}, \theta_t^{x, N})$ such that the state $|\psi_{\rm DQS}\rangle$ is as close as possible to the desired time evolution of the target Hamiltonian $\ham$ at a specified time $\tau$:
$|\psi_{\rm target}\rangle = e^{-i \ham \tau} |\psi_0 \rangle \, .$
From now on the number of entangling gates will be fixed to $n=3$.
As we will show, remarkably, these small quantum circuits will be sufficient to reproduce the dynamics of local observables such as for the lattice Schwinger model, see Fig.~\ref{fig:architecture}(b), even for large systems and large times.
\\

\begin{figure*}[t]
\centering
    \includegraphics[width=1\textwidth]{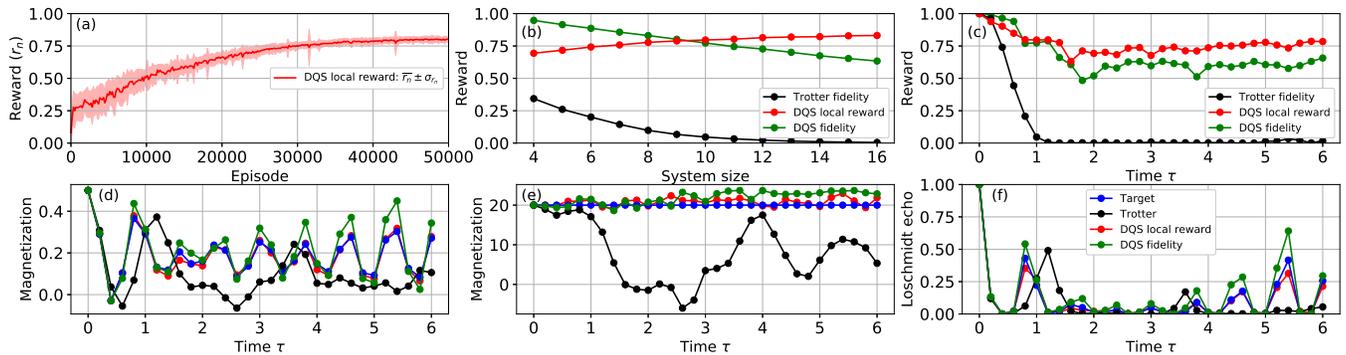}
    \caption{\small Results for the DQS of the LRI model with three entangling gates.
    (a) Evolution of the local reward during training for 100 independent runs for a 16-qubit system (average and standard deviation).
    (b) Rewards of the DQS as a function of the system size (for $\tau =1.0$), and the corresponding Trotter fidelity.
    (c) Rewards of the DQS as a function of time (for a 10-qubit system), and the corresponding Trotter fidelity.
    (d) Magnetization obtained as function of time optimizing for the fidelity (green) and the local reward (red), the corresponding results using Trotterization (black), and the exact time evolution (blue).
    (e) Same as (d) for the energy.
    (f) Same as (d) for the Loschmidt echo.
    } 
    
	\label{fig:LRI}
\end{figure*}

\paragraph{Method.}
We use reinforcement learning (RL) in order to solve this difficult optimization problem.
RL is a subfield of machine learning in which a software agent learns by interacting with an environment and adapting its behavior accordingly. The agent generates sequences of actions in the environment and learns to perform a given task by maximizing a cumulative reward function. 
%
%
RL has seen a recent surge of applications in the field of quantum control for few-body problems \cite{chen2013fidelity,albarran2018measurement,zhang2018automatic,fosel2018reinforcement,bukov2018reinforcement,niu2019universal,yao2020policy,zhang2020topological} as it suits well optimiziation problems consisting of successive actions on a state with high dimensionality. Here, we are interested in the dynamics of quantum many-body problems which is a far more challenging problem.

In this work, we use a modified version of a deep Q-network algorithm \cite{mnih2015human}, a variant of the original Watkins off-policy Q-learning algorithm using artificial neural networks as function approximators \cite{watkins1992q, sutton2018reinforcement}.
While we now summarize the central aspects of the algorithm, further details can be found in Refs.~\cite{SM,sutton2018reinforcement}.

The optimization problem is defined as an episodic RL problem: each episode is divided into a finite number of steps $t=0, \dots, n$, corresponding to the steps of the DQS. At $t=0$, the quantum wave function is in a given initial state $|\psi_0\rangle$. Then, at each step $t$ the agent chooses an action $a_t = (\theta^{xx}_t, \theta_t^{z, 1}, \theta_t^{x, 1}, \dots, \theta_t^{z, N}, \theta_t^{x, N})$ defining the unitary $U_t$ in Eq.~\eqref{eq:action_step}.
  After each action $a_t$, the agent receives a scalar reward $r_t$. At the end of the episode, $t=n$, the reward characterizes how close the final state $|\psi_{\rm DQS}\rangle$ is to the target state $|\psi_{\rm target} \rangle = e^{-i\ham \tau} |\psi_0 \rangle $.
  For intermediate steps, the reward is set to 0 as we do not constraint the specific evolution of the quantum wave function between the initial and target state. 
A representative learning process as a function of episodes is shown in Fig.~\ref{fig:LRI}(a) for the Ising model in Eq.~(\ref{eq:LRI}).

In Q-learning, the goal is to numerically compute the so-called action-value function, $Q(s, a)$ (represented by a neural network in our algorithm), which is defined as the expected total return $\sum_{t=1}^{n} r_t$, given that the environment is in the state $s$ and that the agent takes the action $a$ and acts optimally afterwards.
At each step, the agent chooses an action using current knowledge of the $Q$ functions and uses the obtained feedback to update $Q$ following the Bellman optimality equation \cite{sutton2018reinforcement}.
Importantly, the actions $a$ take continuous values in our case, which is not standard for Q-learning. We have modified our algorithm accordingly so that the $a_t = {\rm argmax}_a Q(s_t, a)$ operation is done by maximizing the output of the neural network with respect to part of its input ~\cite{SM}. 
\\

\paragraph{Reward.}
A central quantity in the RL optimization problem is the cost function measuring the reward and therefore quantifying how close $|\psi_{\rm DQS}\rangle$ is to the target state $|\psi_{\rm target}\rangle$.
The fidelity $|\langle \psi_{\rm DQS} |\psi_{\rm target}\rangle |^2$ is commonly used to compare the two states globally. 
With a limited number of entangling gates, we find, however, that it is challenging to obtain high fidelities for large systems sizes or times.
As a consequence, we now introduce an alternative reward in our RL algorithm, which takes into account that in quantum simulation we are not so much interested in the global many-body wave function but rather in reproducing local observables and correlation functions.
Let $\rho = |\psi_{\rm target}\rangle \langle \psi_{\rm target}|$ and  $\sigma = |\psi_{\rm DQS}\rangle \langle \psi_{\rm DQS}|$ denote the density matrices corresponding to the target and the DQS state.
We then define a local reward 
\begin{align}
    \label{eq:local_reward}
    R_{\rm local} &= 1 - \frac{2}{N(N-1)} \sum_{j<k} \sqrt{D(\rho^{jk} || \sigma^{jk})}
\end{align}
measuring the closeness of reduced density matrices $\rho^{jk}$ and $\sigma^{jk}$ of the subsystem made of sites $j$ and $k$ for $\rho$ and $\sigma$, respectively.
Here $D(\rho || \sigma) = {\rm Tr}\, \rho (\log\rho - \log \sigma)$ is the relative entropy
 and $N$ denotes the number of qubits.
A reward of $R_{\rm local}=1$ means that all $\rho^{ij} = \sigma^{ij}$ and therefore all expectation values and correlation functions are reproduced exactly.
It is a crucial observation that a high local reward $R_{\rm local} = 1 - \epsilon $ can be directly translated into a high accuracy for local observables and correlations functions. For a two-body operator $O = \frac{2}{N(N-1)} \sum_{j<k} O^{jk}$ we have
\begin{align}
  |\langle O \rangle_{\rm target} - \langle O \rangle_{\rm DQS}| \le  \sqrt{2}\, \max_{j,k} \norm{O^{jk}}_{\infty} \cdot  \epsilon,
  \label{eq:errorbound}
\end{align}
where $\norm{\cdot}_{\infty}$ denotes the operator norm. This can be derived using H\"older's inequality for Schatten norms and Pinsker inequality \cite{pinsker1964information, carlen2014remainder}: $|{\rm Tr} \left[(\rho^{jk} - \sigma^{jk}) O^{jk}\right ]| \le \norm{\rho^{jk} - \sigma^{jk}}_1 \norm{O^{jk}}_{\infty} \le \sqrt{2 D(\rho^{jk} || \sigma^{jk})}\  \norm{O^{jk}}_{\infty}$. Similarly, for a single-body operator $O = \frac{1}{N}\sum_j O^j$ we have $|\langle O \rangle_{\rm target} - \langle O \rangle_{\rm DQS}| \le  \sqrt{2} \max_{j} \norm{O^j}_{\infty} \, \epsilon$.

\begin{figure*}[t!]
\centering
    \includegraphics[width=1\textwidth]{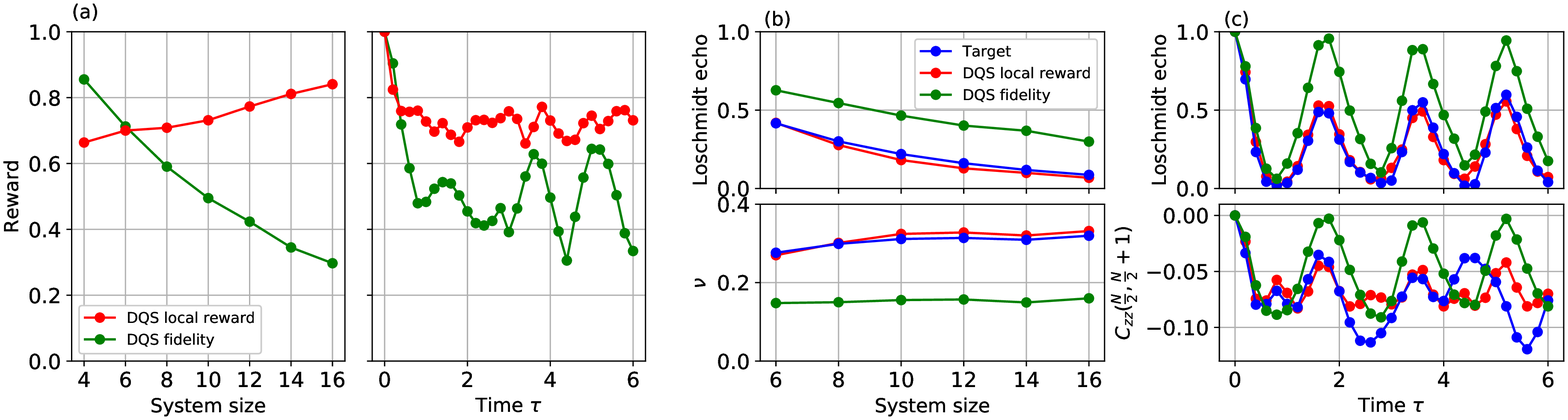}
    \caption{\small Results for DQS of the Schwinger model with three entangling gates.
    (a) Final reward of the DQSs using the fidelity and the local reward~\eqref{eq:local_reward} as a function of system size (for $\tau = 4.0$) and time (for a 10-site system).
    (b) Loschmidt echo and particle number density $\nu$ as a function of system size (for $\tau = 4.0$).
    (c) Loschmidt echo and nearest-neighbor quantum correlations in the middle of the chain of the DQS as a function of time using the fidelity (green) and the local reward Eq.~\eqref{eq:local_reward} (red), and the exact time evolution (blue).}
    
	\label{fig:Schwinger}
\end{figure*}

\paragraph{Results.}
As a first proof of concept, we apply our method to the long-range Ising model 
\begin{equation}
\label{eq:LRI}
  \ham_{\rm LRI} = J \sum_{j<k} \frac{1}{|k - j|^{\alpha}}\sigma^x_j \sigma^x_k + m_x \sum_j \sigma_j^x + m_z \sum_j \sigma_j^z \, ,
\end{equation}
For this system we can directly compare the performance of our approach to a conventional Trotterization procedure, as there exists a straightforward and natural decomposition of the Hamiltonian into the universal set of quantum gates in Eq.~\eqref{eq:quantum_gates} upon choosing $\theta^{xx}_n = J \tau/n$, $\theta^{z, j}_n = m_z \tau/n$, and $\theta^{x, j}_n = m_x \tau/n$.
For concreteness, we will  consider $J=1$ and $m_x = m_z = 2$, and $\alpha = 3$ starting from a fully polarized state $|\psi_0\rangle = |\uparrow\dots \uparrow \rangle$. Let us emphasize, however, that we obtain similar results also for other choices of system parameters. As mentioned before, we fix the number of entangling gates in our circuits to three ($n=3$) both for the Trotterized dynamics as well as the circuit from RL. 

The learning of the agent is witnessed by looking at the evolution of the reward as a function of episodes, i.e., during the learning process, shown in Fig~\ref{fig:LRI} (a) (for $16$ qubits at $\tau = 1$ using the local reward). Starting from the Trotterized circuit, the agent progressively improves the circuit until convergence. The mean value of the maximum rewards throughout each independent run
 is shown in Fig.~\ref{fig:LRI} (b) upon varying the system size $N$ (at fixed $\tau=1.0$). As opposed to the Trotter fidelity, which decays exponentially, the DQS rewards remain at large values. The obtained fidelity decays with the system size $N$ but only linearly at the considered $N$, and the local reward is remarkably unaffected. Now if we fix the system size and increase $\tau$, the Trotterization also fails eventually. In Fig.~\ref{fig:LRI} (c) we show this (for $10$ qubits) together with the DQS results for both types of rewards. To give a more physical perspective to our results, we also compare the values of physical observables resulting from DQS, Trotterization, and from the actual dynamics (using exact diagonalization). Figures~\ref{fig:LRI} (d), (e) and (f) show the magnetization, the energy, and the Loschmidt echo $|\langle \psi_0 | \psi \rangle|^2$.
For the 10-qubit system, after $\tau=1$ it is clear that the system enters a regime where the Trotterization with $n=3$ fails. At the same time, there is a drop in performance of our algorithm, but the reward converges to a finite value as $\tau$ increases. 
Importantly, when translated in terms of physical observables the resulting quantum circuits are much more successful than Trotterization. The magnetization, energy, and Loschmidt echo are well reproduced by the DQS, especially when the local reward is used. This indicates that our algorithm can systematically find a circuit bringing the initial state to an arbitrary target state (e.g. the time evolved state with arbitrary large time) using only three entangling gates.
\\

Having demonstrated that our RL based method with local reward exhibits a remarkable performance for the Ising model, we now aim to go one step ahead by studying a system where no natural decomposition into a Trotter sequence exists for the considered universal set of quantum gates.
For that purpose we focus in the following on the lattice Schwinger model:
\begin{align}
  \ham_{\rm S} &= w \sum_{j} \left[ \sigma_j^{+} \sigma_{j+1}^{-} + {\rm H.c.} \right]+ \frac{m}{2} \sum_j (-1)^n\sigma_n^z  \nonumber \\
   & \quad + \frac{J}{2} \sum_{j=1}^{N-1} \left[  \sum_{m=1}^j\left[\sigma^z_m + (-1)^m \right] \right]^2.
\label{eq:schwinger}
\end{align}
which is represented here in the Kogut-Susskind Hamiltonian formulation \cite{schwinger1962gauge,kogut1975hamiltonian}, as it has been recently realized experimentally using DQS based on Trotterization~\cite{martinez2016real}.
Concerning the non-equilibrium protocol we closely follow the experiment~\cite{martinez2016real}. We start from the N\'eel state (corresponding the bare vacuum) and then apply 
$e^{-i \ham_{\rm S} \tau}$
with $w = J = 1$ and $m=0.5$. Further, we use $\alpha=1$ for the entangling gates in the DQS in Eq.~\eqref{eq:quantum_gates}, as this represents one of the optimal working points in systems of trapped ions.

Even more so than with the LRI model, optimizing with the fidelity only results in suboptimal sets of parameters as can be seen in Fig.~\ref{fig:Schwinger}. Both short-range and long-range couplings are present in the lattice Schwinger model, and thus reproducing the dynamics with only three all-to-all entangling gates is particularly challenging. Nevertheless, we show that  better sets of parameters do exist and are obtained when using the local reward. Interestingly, as for the LRI model, the performance of the algorithm with the local reward does not plummet as the system size increase, and physical observables are significantly better reproduced with the local reward than with the fidelity, as shown Fig.~\ref{fig:Schwinger} (b) and (c) and in Fig.~\ref{fig:architecture} (b), where the particle number density $\nu = \frac{1}{2N}\sum_{j=1}^{N} \langle (-1)^j \sigma^z_j + 1 \rangle$ is shown, which has also been measured in the recent experiment~\cite{martinez2016real}.

While $\nu$ as a few-body operator is directly covered by the local reward, the Loschmidt echo is a global quantity, but can be nevertheless reproduced remarkably well.
To explore further the performance of our RL approach can, we compare in Fig.~\ref{fig:Schwinger} (c) the obtained dynamics for a two-body quantum correlation function against the exact solution.  There we show results for the connected correlator $C_{zz}(N/2, N/2+1)$ in the middle of the chain where $C_{zz}(j, j+1) = \langle \sigma^z_j \sigma^z_{j+1} \rangle - \langle \sigma^z_j\rangle \langle \sigma^z_{j+1} \rangle$. While the two-body operator seems not as well reproduced as the single-body operator for long times, this is different for $\tau \lesssim 2.0$ when using the local reward. This is remarkable as the overall signal strength of $C_{zz}(N/2, N/2+1)$ is much smaller than what one would expect on the basis of the bound in Eq.~(\ref{eq:errorbound}).
\\

\paragraph{Outlook.}

For the considered problems 3 entangling gates have turned out to be typically sufficient for an accurate DQS of local observables, remarkably. In the future it might be important to increase the number of gates for higher precision, where convergence of our algorithm turns out to become progressively challenging. This might be remedied for instance by either utilizing more advanced neural network structures, e.g., recurrent neural networks or long short-term memories, or by reducing the number of independent variational parameters in the optimization problem using physical insights, in particular, by utilizing symmetries.

The current scheme requires an exact theoretically known reference of the target state, which we obtain using exact diagonalization. The overarching goal of DQS, however, is to address scenarios which are beyond such a theoretical description and therefore without such exact reference available. For the future it might be a key open question whether it is possible to obtain optimized time-evolution operator factorizations using reinforcement learning without an exact reference. However, for current typical DQS scenarios such a regime of quantum supremacy is not yet reached, so that our algorithm represents a central contribution to push DQS significantly beyond what has been achieved up to now in terms of system size and simulation time.
\paragraph{Acknowledgment.}
We acknowledge Peter Zoller, Rick van Bijnen, and Christian Kokail for the fruitful discussions. 
This project has received funding from the European Research Council (ERC) under the European Unions Horizon 2020 research and innovation programme(grant agreement No. 853443), and M. H. further acknowledges support by the Deutsche Forschungsgemeinschaft via the Gottfried Wilhelm Leibniz Prize program.

\bibliographystyle{apsrev4-1}
%

\pagebreak
\widetext
\begin{center}
\textbf{\large Supplemental Material}
\end{center}
\setcounter{equation}{0}
\setcounter{figure}{0}
\setcounter{table}{0}
\setcounter{page}{1}
\makeatletter
\renewcommand{\theequation}{S\arabic{equation}}
\renewcommand{\thefigure}{S\arabic{figure}}
\renewcommand{\bibnumfmt}[1]{[S#1]}

\section{Deep $Q$-Learning}
$Q$-learning is a reinforcement learning (RL) algorithm to teach an agent what action to take in an environment under what circumstances. It is model-free (does not need require a model of the environment).
For a finite Markov decision process, Q-learning finds a policy that maximizes the expectation value of the cumulative rewards: the reward over all successive steps starting from the current state.

Formally, RL involves a set of states $S$ and a set of actions $A$. By taking an action $a_t \in A$ at step $t$, the agent makes the environment transition from a state $s_t$ to another state $s_{t+1}$. Performing an action $a_t$ in a given state $s_t$ provides the agent with a reward $r_t$ (a scalar). For Markov decision processes, the state $s_{t+1}$ and the rewards $r_{t}$ only depend on the action $a_t$ and state $s_t$, and not on the previous ones. The episodic RL problem can be represented by a trajectory
 \begin{equation}
 	s_0 \rightarrow a_0 \rightarrow r_0 \rightarrow s_1 \rightarrow  \dots \rightarrow a_{n-1} \rightarrow r_{n-1} \rightarrow s_{n},
 \end{equation}
 where $n$ is the number of steps in an episode.
 As explained in the main text, $t=0, \dots, n$, corresponding to the steps of the DQS. At $t=0$, the quantum wave function is in a given initial state $|\psi_0\rangle$ and $n$ is the number of entangling gates in the DQS. The actions $a_t = (\theta^{xx}_t, \theta_t^{z, 1}, \theta_t^{x, 1}, \dots, \theta_t^{z, N}, \theta_t^{x, N}) \in A$ define unitary operators $U_t$ [see Eq.~\eqref{eq:action_step}], and the states $s_t$ represent the quantum wave functions of the qubits $|\psi_t\rangle = U_t \cdots U_2 U_1 |\psi_0\rangle$. However, we internally represent the state as $[t, U_t]$. It has the advantage of having a fixed dimension, and even though it does not contain the full information of the wave function, the efficiency of the algorithm the RL algorithm is not affected because of the small number of steps in the episode. Finally, the reward $r_t \in [0, 1]$ is chosen as

\begin{align}
  r_t = \left\{
  \begin{aligned}
  	0\,&, \quad {\rm if}\, t < n  \\
  	f(|\psi_{\rm target}\rangle , |\psi_{\rm DQS} \rangle)\,&, \quad {\rm if}\, t = n.
  \end{aligned}
  \right. ,
\end{align}
where the function $f$ is either the fidelity reward of the local reward defined in the main text. At the end of the episode, $t=n$, the reward characterizes how close the final state $|\psi_{\rm DQS}\rangle$ is to the target state $|\psi_{\rm target} \rangle = e^{-i\ham \tau} |\psi_0 \rangle $. For intermediate steps, the rewards is set to 0 as we do not constraint the specific evolution of the quantum wave function between the initial and target state.

An essential part of any RL algorithm is to define a policy -- how to choose the next action $a_t$ given that the environment is in the state $s_t$. 
 The central objects in $Q$-learning are the action-value functions, or $Q$ functions, which are used to define the policy. The optimal action-value function $q_*(s, a)$ is defined as the expected total return $\sum_{t=1}^{n} r_t$, given that the environment is in the state $s$ and that the agent takes the action $a$ and acts optimally afterwards. Once the $q_*$ is known, the optimization problem becomes trivial, and the optimal policy is $a_t = {\rm argmax}_a q_*(s_t, a)$. The goal of Q-learning is to build an approximation $Q(s, a)$ through clever exploration of the environment. The optimal $Q$ function satisfies the Bellman optimality equation \cite{sutton2018reinforcement}, which is solved numerically using temporal difference learning through the update rule

\begin{equation}
\label{eq:update}
  Q(s_t, a_t) \leftarrow Q(s_t, a_t) + \alpha[r_t + \max_a Q(s_{t+1}, a) - Q(s_t, a_t)],
\end{equation}
where $\alpha$ is a learning rate. In Q-learning the actions are chosen using current estimations of the $Q$ function, $a_t = {\rm argmax}_a Q(s_t, a)$, on top of which noise is added in order to increase exploration of the environment (the algorithm is thus off-policy as the behavior policy differs from the target policy). After meaning episodes, the algorithm converges and the policy obtained using $Q(s, a)$ (i.e. $a_t = {\rm argmax}_a Q(s_t, a)$) results in near-optimal sequences of gates.

In the vanilla $Q$-learning algorithm, $s$ and $a$ only take a finite number of discrete values and $Q(s, a)$ is a matrix. To cope with complex environements (with continuous-valued states or highly-dimensional states, or both), function approximators are used for $Q$.

In Deep $Q$-learning, the $Q$ function is approximated with a neural network. For instance, in the original papers of DeepMind \cite{mnih2013playing,mnih2015human}, convolutional networks are used to process the state of the environment made of pixels on a screen. In our algorithm, we use three-layer dense neural networks because the dimensionality is not so high.

Typically, the neural network has a state $s$ as an input, and outputs different values $Q(s, a)$ for all the possible discrete values of $a$.
In our problem, the actions themselves are also continuous. In this case, there are Actor-Critic RL algorithms that solve the continuous action problem, such as deterministic policy gradient \cite{lillicrap2015continuous}. We tried using deterministic policy gradient, but it turned out that a more simple modification of the deep $Q$-network worked better. The neural networks used to represent $Q(s, a)$ have both actions $a$ and states $s$ as inputs, and a single scalar as output for the value of $Q(s, a)$. The only non-trivial difference from the usual deep $Q$ network is how to calculate ${\rm argmax}_a Q(s, a)$: we use gradient ascent (with Nesterov accelerated gradient) to maximize $Q(s, a)$ with respect to the action inputs and with fixed state inputs.

In addition, RL is known to be unstable or divergent when a nonlinear function approximators such as a neural networks are used \cite{mnih2015human}.
The instability comes from the correlations present in the sequence of data used to train the neural network, the fact that small updates to $Q$ can significantly change the policy, and the correlations between $Q$ and the target values. Indeed, the $Q$ function is theoretically both used to choose the behavior $a_t$ and the target in Eq.~\eqref{eq:update}. We use common techniques developed to solve these problems: 
experience replay, and iterative updates of the target $Q$ network (a separate network only periodically updated in order to reduce correlations with the target)

In the following, we present the details of the algorithm used with pseudocode. The hyperparameters are given as we used them after tuning (through grid search).

\begin{algorithm}[H]
    \centering
    \caption{Deep Q Networks}\label{algorithm}
    \begin{algorithmic}[1]
    	\State $N \gets$ number of sites, $n \gets$ number of steps (typically $n=3$)
    	\State Initialize replay memory $\mathcal{D}$ with capacity of $50$ episodes.
    	\State $d_a \gets 2 N + 1$ (dimension of actions)
    	\State Initialize $Q_{\rm behavior}$ with random weights (dense NN with architecture [150: tanh, 40: relu, 1: $\sigma$] and input dim. $n + 2d_a$).
    	\State $Q_{\rm target} \gets Q_{\rm behavior}$ (copy the NN weights)
    	\For{episode $=1$ to $5 \cdot 10^4$}
    		\State Set value of $\epsilon$ for exploration according to some schedule (from $1$ to $0.005$ exponentially)
    		\State $s_0 \gets [onehot(t=0)] + [0]*d_a$
    		\State every 
    		\For{$t=0$ to $n-1$}
    		\State Select action $a_t \gets \Call{choose\_action}{s_t}$.
    		\State Add Gaussian noise with standard deviation $\epsilon/2$ to $a_t$.
    		\State $r_t \gets 0$, $s_{t+1} \gets [onehot(t), a_t]$
    	\EndFor
    	\State $r_{n-1} \gets \Call{Reward}{a_0, \dots, a_{n-1}}$, scalar reward calculated from the target and final states (fidelity or local reward).

    	\State Push all ($s_t, a_t, r_t, s_{t+1}$) transitions of the episode to memory $\mathcal{D}$.
    	\For{$(s_t, a_t, r_t, s_{t+1})$ in $\mathcal{D}$}
    		\State set target $y \gets r_t + \max_a Q_{\rm target}(s_{t+1}, a)$
    		\State Perform backpropagation on $Q_{\rm behavior}(s_t, a_t)$ with target $y$ (with adam optimizer and logcosh loss)
    	\EndFor
        \EndFor
        \State \textbf{Return} sequence ($a_0, \dots, a_{n-1}$) with max reward $r_{n-1}$.
        \\
		\Function{choose\_action}{$s_t$}
			\State Find ${\rm argmax}_a Q_{\rm behavior}(s_t, a)$ by performing gradient ascent (with Nesterov accelerated gradient) on $Q_{\rm behavior}(s_t, \cdot)$.
			\State Start from $15$ different $a$ (with all components between $-1$ and $1$). momentum $= 0.9$, learning rate $=0.6$.
			\State Return best $a$.
		\EndFunction
		\\
		\Function{reward}{$a_0, \dots, a_{n-1}$}
			\For{$t=0$ to $n-1$}
				\State Obtain physical parameters by rescaling the components of $a_t$:
				\State $\theta^{xx}_t \gets a_t[0] \cdot 0.2$
				\State $(\theta^{z, 1}_t ,\dots, \theta^{x, N}_t) \gets a_t[1:-1] \cdot 0.4$
			\EndFor
			\State Compute $|\psi_{\rm DQS}\rangle = U_{n} \cdots U_2 U_1 |\psi_0\rangle$
			\State Return reward calculated from $|\psi_{\rm DQS}\rangle$ and $|\psi_{\rm target}\rangle$.
		\EndFunction
    \end{algorithmic}
\end{algorithm}

\begin{figure}
\centering
\includegraphics[width=0.6\textwidth]{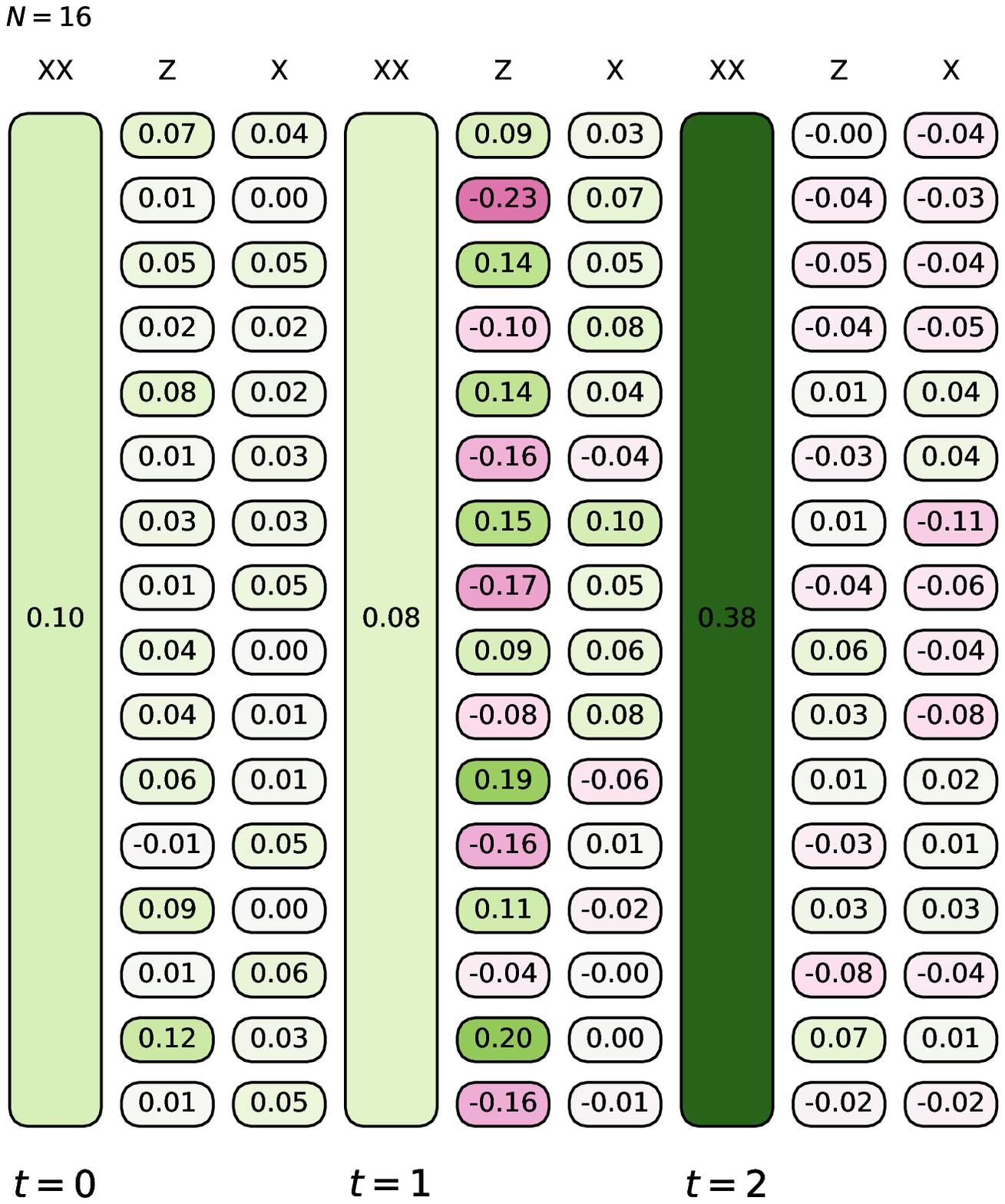}
    \caption{\small Example of a resulting DQS quantum circuit obtained from our algorithm (16-qubit DQS using local reward for the Schwinger model).}

	\label{fig:gates_drawing}
\end{figure}

\end{document}